\documentclass[%
 amsmath,amssymb,
 reprint,
]{revtex4-1}

\usepackage{graphicx}
\usepackage{dcolumn}
\usepackage{bm}

\usepackage[utf8]{inputenc}
\usepackage[T1]{fontenc}
\usepackage{mathptmx}

\usepackage{xfrac}
\usepackage[version=3]{mhchem}
\usepackage{balance}
\usepackage{multirow}
\usepackage{xfrac}
\usepackage{placeins}
\usepackage[resetlabels]{multibib}
\newcites{SI}{References}

\begin{document}

\title{Suitability of $\beta$-Mn$_2$V$_2$O$_7$/$\beta$-Cu$_2$V$_2$O$_7$ solid solutions for photocatalytic water-splitting}

\author{Silviya Ninova}
\affiliation{University of Bern, Department of Chemistry and Biochemistry, Freiestrasse 3, 3012 Bern, Switzerland}

\author{Michal Strach}
\altaffiliation[Present address: ]{Chalmers University of Technology, Department of Physics, 412 96 G\"oteborg, Sweden}
\affiliation{EPFL Valais Wallis, EPFL SB ISIC LNCE, Rue de l'Industrie 17, Case postale 440, 1951 Sion, Switzerland}

\author{Raffaella Buonsanti}
\affiliation{EPFL Valais Wallis, EPFL SB ISIC LNCE, Rue de l'Industrie 17, Case postale 440, 1951 Sion, Switzerland}

\author{Ulrich Aschauer}
\email[Corresponding author: ]{ulrich.aschauer@dcb.unibe.ch}
\affiliation{University of Bern, Department of Chemistry and Biochemistry, Freiestrasse 3, 3012 Bern, Switzerland}

\date{\today}

\begin{abstract}
The pyrovanadates $\beta$-Mn$_2$V$_2$O$_7$ and $\beta$-Cu$_2$V$_2$O$_7$ were previously investigated as photoanode materials for water splitting. Neither of them, however, was found to be sufficiently active. In this work we predict the properties of solid solutions of these two structurally similar pyrovanadates via density functional theory calculations to explore the suitability of their band structure for water splitting and to assess their ease of synthesis. We predict that substitution of up to 20\% Cu or Mn into $\beta$-Mn$_2$V$_2$O$_7$ and $\beta$-Mn$_2$V$_2$O$_7$ respectively leads to a narrowing of the band gap, which in the former case is experimentally confirmed by UV-vis spectroscopy. Calculations for solid solutions in the intermediate composition range, however, yield nearly constant band gaps. Moreover, we predict solid-solutions with higher substitution levels to be increasingly difficult to synthesize, implying that solid solutions with low substitution levels are most relevant in terms of band gaps and ease of synthesis.
\end{abstract}

\maketitle


\section{Introduction}
Solar water splitting is a promising strategy to convert solar energy to \ce{H2} fuel. Photoelectrode materials used in water-splitting applications must fulfil strict criteria: they must have band gaps small enough to absorb a large portion of the solar spectrum, their bands edges must provide a sufficient overpotential for the oxygen and hydrogen evolution reactions (OER and HER on the anode and cathode respectively) and the materials need to maintain high (electro)chemical stability under application conditions.

\begin{figure}
	\centering
	\includegraphics[scale=0.95]{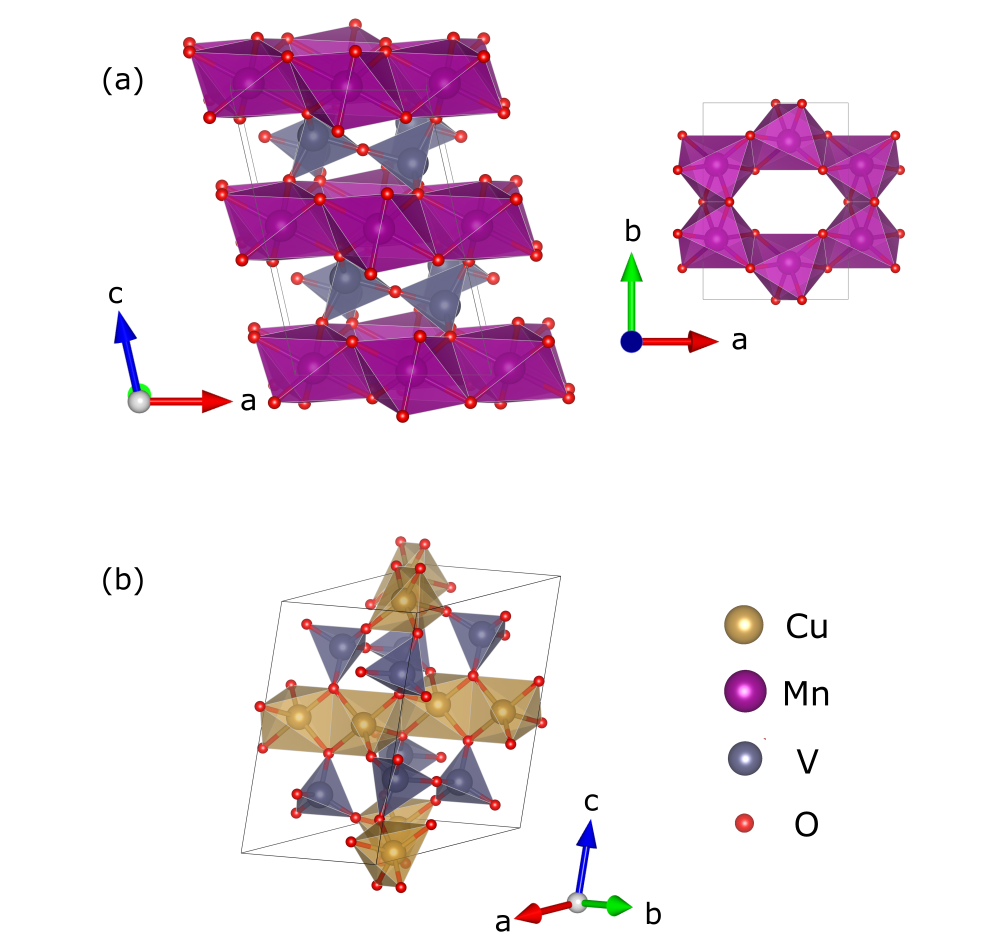}
	\caption{(a) $\beta$-Mn$_2$V$_2$O$_7$ and (b) $\beta$-Cu$_2$V$_2$O$_7$ have a similar crystal structure, the honey-comb geometry of the MnO$_6$-octahedra in the former assuming the form of chains of CuO$_5$-square pyramids in the latter.}
	\label{fig:CMVO-structures}
\end{figure}

The two pyrovanadates, $\beta$-Mn$_2$V$_2$O$_7$ (MVO, Fig. \ref{fig:CMVO-structures}a) and $\beta$-Cu$_2$V$_2$O$_7$ (CVO, Fig. \ref{fig:CMVO-structures}b), were previously investigated as potential photoanodes.~\cite{Yan2015,Guo2015,Newhouse2016,Zhou2015,Jiang2018} MVO has a sub-2\,eV band gap and was found to be stable under illumination in alkaline conditions. It is, however, inactive for the OER without a facile redox couple~\cite{Yan2015}. CVO, on the other hand, is OER active but has a small photoelectrocatalytic activity due to sizeable excitonic effects~\cite{Wiktor2018}, short carrier diffusion length and slow water oxidation kinetics~\cite{Song2020}. In addition, it has a relatively large band gap of 2\,eV~\cite{Zhou2015,Guo2015,Song2020}. The performance of CVO can, however, be enhanced by the addition of OER catalysts~\cite{Guo2015,Newhouse2016,Kim2017}.

While these two ternary pyrovanadates are thus not ideal candidates for photoanodes, improved efficiencies were previously reported for quaternary vanadates ~\cite{Liu2005,Rajeshwar2018,Hassan2018}, that can be considered to be solid solutions of ternary vanadates. Solid-solutions were previously used to engineer photocatalytic activity in other materials, mostly via tuning the band gap ~\cite{Kudo2002, Tsuji2004, Tsuji2005a, Tsuji2005b, Tsuji2006, Xing2006}. In the case of pyrovanadate, they have so far only been investigated for thermoelectric~\cite{Sotojima2007} and negative thermal-expansions applications~\cite{Katayama2018, Wang2019a}. Structures of MVO-CVO solid solutions, in particular, were refined in the low CVO content limit and suggested that Cu addition can stabilize the $\alpha$ phase also above room temperature~\cite{Krasnenko2013Structural}.

In the present work, we predict the crystal structure, electronic properties and phase stability of $\beta$-(Cu$_x$Mn$_{1-x}$)$_2$V$_2$O$_7$ solid solutions using density functional theory calculations and discuss their properties in view of photoanode applications for solar water splitting. We show that Mn-rich solid solutions indeed show a reduced band gap compared to pure MVO but that band gaps stay nearly constant for higher levels of substitution before sharply rising when approaching the CVO composition. For solid solutions with low Cu content, these predictions are supported by experimentally determined band gaps. Given that our calculated formation enthalpies hint at difficult synthesis of compositions far from the end members, solid solutions with low substitution levels are thus most relevant for photoelectrocatalytic water splitting.

\section{Methods}

\subsection{Computational}
We carried out calculations using the Quantum ESPRESSO package~\cite{QE-2009, QE-2017} with the Perdew-Burke-Ernzerhof (PBE) gradient-corrected exchange-correlation functional~\cite{PBE}. Ultrasoft pseudopotentials~\cite{UltraSoftPseudo} are used for all elements. The wavefunctions are expanded in plane waves with a kinetic energy cut-off of 40\,Ry and a cut-off of 320\,Ry for the augmented density. A Hubbard $U$ correction was applied to improve the description of all metal  3\textit{d}-states. We used the linear-response scheme proposed by Cococcioni~\cite{Cococcioni2005} to determine $U$ values of 4.1\,eV for Mn and 3.9\,eV for V, which agree well with other theoretical studies of MVO.~\cite{Medvedeva2015,Yan2015} The same approach resulted in $U$=9.8\,eV for copper, which was combined with $J$=1.2\,eV, motivated by the fact that $J$ improves the description of the magnetic state of CuO, which is also a Cu(II) compound. ~\cite{Himmetoglu2011}  This setup differs from other CVO calculations that used either an effective $U_\mathrm{eff}=U-J=6.52$\,eV~\cite{Yashima2009,Zhou2015} or $U_\mathrm{Cu}$=7\,eV and $J_\mathrm{Cu}$=1\,eV.~\cite{Tsirlin2010} Despite these differences, all methods yield a similar electronic structure. Band gaps and band-edge positions were calculated with the HSE hybrid functional~\cite{HSE, HSE06}. The portion of exact exchange was fixed to 16\%, so as to match the experimentally observed band gaps of both pure compounds (see ESI\dag\ Table S4). 

We used experimental crystal structures as starting points and subsequently relaxed all lattice parameters and internal coordinates within the PBE+$U$ setup. The experimental MVO structure is monoclinic (space group C2/m) with $a=6.713$ \AA, $b=8.725$ \AA, $c=4.969$ \AA\ and $\beta$=103.6$^{\circ}$,~\cite{Liao1996}  while CVO crystalizes in space group C2/c with $a=7.689$ \AA, $b=8.029$ \AA, $c=10.106$ \AA\ and $\beta$=110.3$^{\circ}$.~\cite{Hughes1989} To obtain cells with the same number of atoms, we doubled the MVO cell along the \textit{c} axis, yielding cells with 8 Cu/Mn atoms that allow for 12.5\%, 25\% and 50\% solid solutions. The Brillouin zone was sampled with 6$\times$6$\times$4 Monkhorst-Pack meshes~\cite{MonkhorstPack} in all cases. Geometries were relaxed until forces converged below 0.03\,eV/\r{A} and stressed below 0.5\,kbar, which for MVO and CVO resulted in deviations of 1.8\% and 3.4\% respectively from the experimental volume (see ESI\dag\ Table S1).

The band-edge positions with respect to the normal hydrogen electrode (NHE) as a function of the copper content $x$ were computed using the procedure outlined by Butler and Ginley~\cite{Butler1978} and Xu and Schoonen~\cite{Xu2000}, typically used in high-throughput calculations~\cite{Castelli2012}. The valence and conduction band edges are given by
\begin{equation}
	E_{VB,CB}=E_0+(\chi_{Cu}^{2x}\chi_{Mn}^{2-2x}\chi_V^{2}\chi_O^{7})^{1/11} \pm E_{gap}/2,
\end{equation}
where $E_0=-4.5eV$ is the difference between the NHE and vacuum, and $\chi$ is the Mulliken electronegativity of each atom in the solid solution.

\subsection{Experimental}

The samples for the UV-vis characterization were prepared as it follows. Powders of copper (I) acetate (Cu(OAc) 97\% Sigma-Aldrich), manganese (II) acetate (Mn(OAc)$_2$ 98\%, Sigma-Aldrich), vanadium acetylacetonate (VO(acac)$_2$ 98\% Sigma-Aldrich) in ratios according to the desired stoichiometry of the final product were dissolved in methanol at 70$^{\circ}$C. The as-prepared solutions were deposited on quartz substrate by dip coating (rate 5\,cm/min, 60\,s drying, repeated for 60 cycles). The samples for X-Ray diffraction (XRD) were prepared similarly but annealed as powders for sufficient data analysis. Details on the UV-Vis and XRD experiments can be found in Ref ~\cite{Wiktor2018}.

\section{Results and Discussion}

The pure pyrovanadates MVO and CVO share a similar crystal structure with layers of edge-sharing Mn/Cu polyhedra  separated by V$_2$O$_7^{4-}$ layers (see Fig. \ref{fig:CMVO-structures}), albeit in different space groups. The MnO$_6$ octahedra in MVO form a honey-comb structure, while due to the Jahn-Teller effect caused by the Cu(II) d$^9$ electron configuration, CVO adopts a structure with chains of CuO$_5$ square pyramids. The Cu--O bonds in the calculated CVO structure are shorter (1.98-2.21\,\AA) compared to the Mn--O bonds in MVO (2.13-2.28\,\AA), leading to different V--O--V angles of 142.9$^{\circ}$ in CVO and 179.6\,$^{\circ}$ in MVO respectively. Moreover the V$_2$O$_7$ layers are buckled, leading to alternating short and long metal-metal distances along the $c$-axis that are 3.36 and 3.50\,\AA\ in MVO (see ESI\dag\ Fig. S2), as well as 3.05 and 3.22\,\AA\ in CVO (see ESI\dag\ Fig. S3). These computed structural properties agree well with the available experiments~\cite{Liao1996,Hughes1989}.

In terms of their magnetic properties, the two pyrovanadates differ. MVO is ferromagnetic (FM) in our 0\,K DFT calculations, with a high-spin Mn$^{2+}$ configuration, the lowest antiferromagnetic (AFM) state lying 4.6\,meV per Mn atom higher in energy, which is in agreement with existing studies.~\cite{Yan2015} CVO in contrast is AFM, the FM state lying 2.5\,meV per Cu atom higher in energy, which is also in agreement with existing studies.~\cite{Touaiher2004,Yashima2009,Tsirlin2010} These lowest energy magnetic states will be kept throughout the study of the solid solutions, where different magnetic arrangements were tested only on the substituted ions.

\begin{figure}
	\centering
	\includegraphics[scale=1.0]{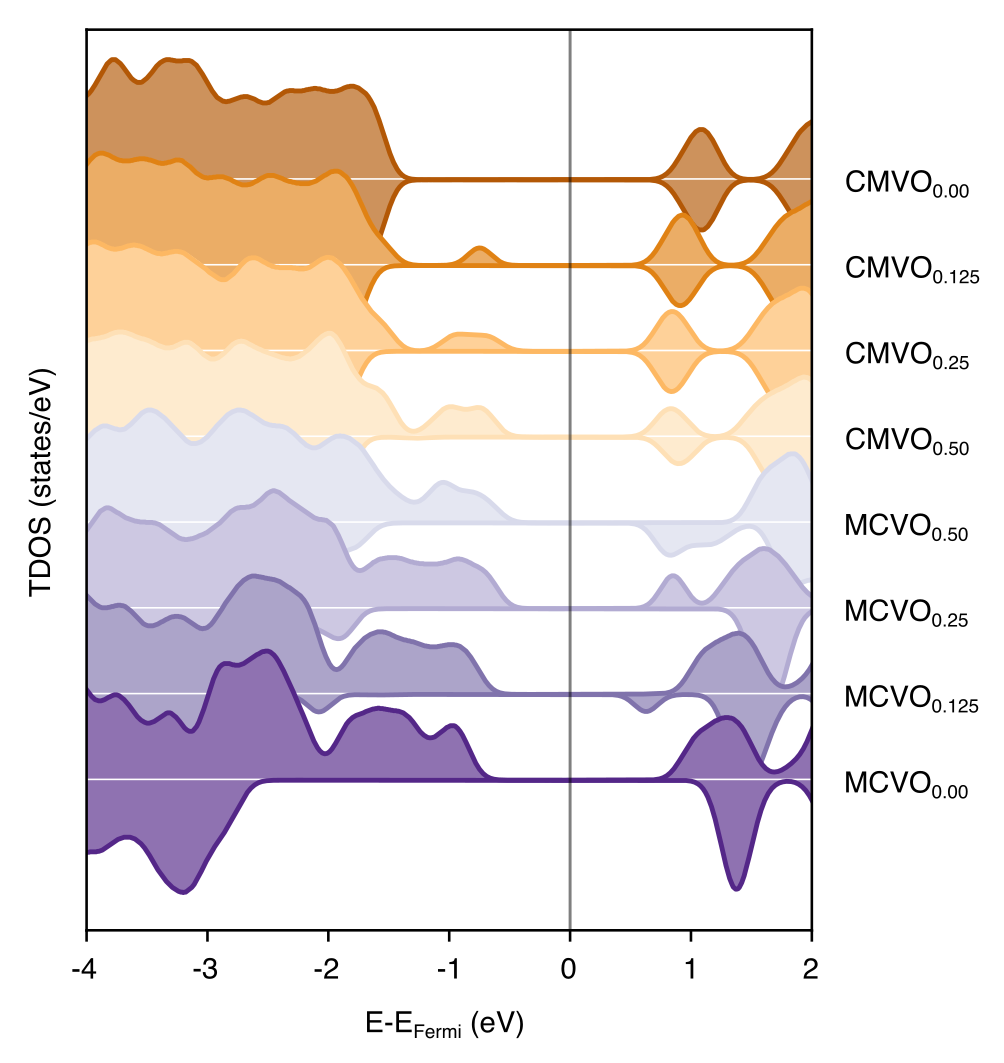}
	\caption{Total density of states (TDOS) of the solid solutions obtained with the HSE functional as a function of their chemical composition. A Gaussian broadening of 0.01\,eV was used.}
	\label{fig:CVO-MVO-all}
\end{figure} 

The solid-solution models were built from the two pure compounds, MVO and CVO, given their slightly different space group. In each case we substituted $\sfrac{1}{8}$, $\sfrac{1}{4}$ and $\sfrac{1}{2}$ of the parent cation and designate the solid solutions $\beta$-(Mn$_{1-x}$Cu$_x$)$_2$V$_2$O$_7$ as MCVO$_x$ and $\beta$-(Cu$_{1-x}$Mn$_x$)$_2$V$_2$O$_7$ as CMVO$_x$, where $x$ is the substitution level (0.125, 0.250, 0.500). Such a metal exchange affects the crystal structures, due to the different crystal radii of the two elements - 0.97\,\AA\ for Mn and 0.79\,\AA\ for Cu~\cite{Shannon1976}. The Cu substitution for instance leads to shorter bonds with oxygen atoms in the $bc$-plane, while those along $a$ expand due to the Jahn-Teller effect. This is also reflected by changes in lattice constants (see ESI\dag\ Table S1) and a reduction of the V--O--V angle from 179.6 to 167.2$^{\circ}$ for instance in MCVO$_{0.25}$. For substitution levels higher than $x=0.125$, the Cu ions tend to group together (see ESI\dag\ Fig. S2), such clustered geometries being 0.1 and 0.04\,eV lower in energy in MCVO$_{0.25}$ and MCVO$_{0.50}$ respectively (see ESI\dag\ Table S2). This can be interpreted as a less strained way of locally accommodating chains of Cu square pyramids. While the Cu spins seem to favour a FM alignment, the FM energies are very close to those of the AFM arrangement (see ESI\dag\ Table S2) and we will use the lowest energy in each case for further analysis. Mn ions substituted into CVO, on the other hand, show a slight preference to occupy short-couple metal sites, to yield short metal-metal separation in both CMVO$_{0.25}$ and CMVO$_{0.50}$ (see ESI\dag\ Fig. S3 and ESI\dag\ Table S3). All Mn substitution in CVO result in energetically similar FM and AFM states with a slight preference for the former, which we will consider for subsequent analysis. It is important to note that CMVO$_{0.50}$ and MCVO$_{0.50}$ models have different geometries, the former being 0.48\,eV lower in energy. This implies that there is no unique intermediate structure, but the exact transformation mechanism based on metal diffusion is beyond the scope of the present work.

The above computational predictions agree with the fact that experimentally significant structure changes are observed in the XRD pattern already with addition of 10\% Cu (see ESI\dag\ Fig. S1). While reference patters cannot be used to reliably identify a structure in such complex systems, no apparent secondary phases were identified and we assume full incorporation of Cu into the MVO lattice at low doping levels.

Important changes also occur in the electronic structure as a function of the solid solution's composition. The pure compounds have very characteristic densities of states: The valence region of MVO contains Mn 3\textit{d} and O 2\textit{p} states, whereas in the conduction band there are mainly V 3\textit{d} states (see ESI\dag\ Figs. S5 and S7). By contrast, the band gap of CVO separates occupied O 2\textit{p} and empty Cu 3\textit{d} states. The gradual substitution of Cu with Mn entails the appearance of occupied peaks of Mn and O states above the valence band, which merge with the valence band for higher substitution levels (see Fig. \ref{fig:CVO-MVO-all}). In the conduction region, on the other hand, the typical peaks of Cu 3\textit{d} states just above the band gap gradually give way to the vanadium states as the copper content decreases.

\begin{figure}
	\centering
	\includegraphics[scale=1.0]{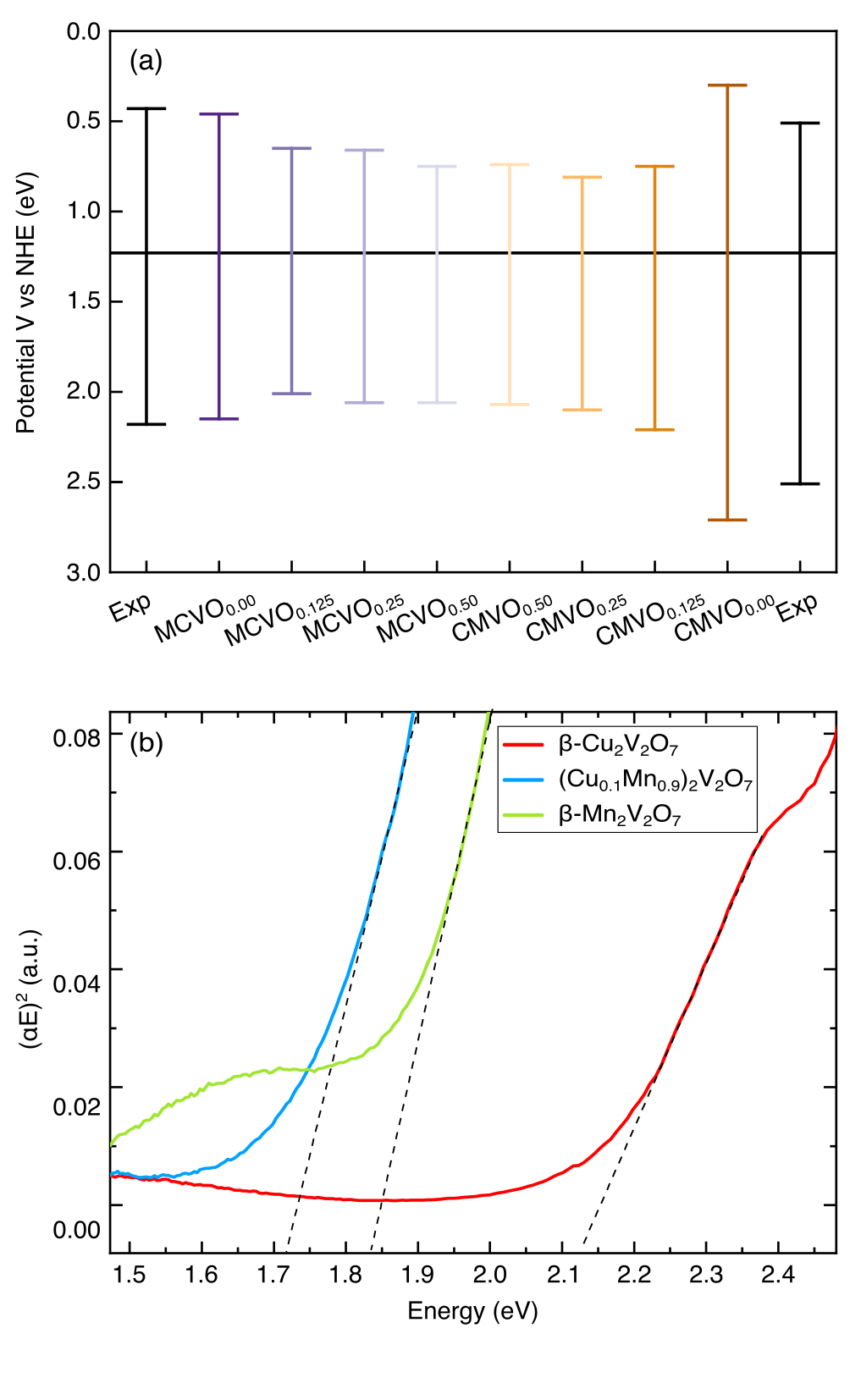}
	\caption{(a) HSE band-edge positions and (b) Tauc plots from UV-vis experiments assuming direct band gap.}
	\label{fig:CVO-MVO-bandgap-qe}
\end{figure}

These changes in the electronic structure are also reflected in the resulting band gap, which we extract from HSE calculations based on DFT+$U$ structures. For solid solutions with up to around 20\% of Mn in CVO we observe a pronounced narrowing of the band gap (see Fig. \ref{fig:CVO-MVO-bandgap-qe}a and ESI\dag\ Fig. S4). Then the band gap remains almost constant before rising sharply towards the CVO composition. These theoretical predictions are supported by experimental UV-vis spectroscopy (see Fig. \ref{fig:CVO-MVO-bandgap-qe}b) that points to a band gap reduction slightly larger then 0.1 eV going from MVO to MCVO$_{0.1}$ followed by a significantly larger band gap of over 2\,eV for CVO. These results suggest that MVO could benefit from a $\approx 10$\% Cu substitution to absorb a larger fraction of solar light, while CVO will have to be substituted with $>20$\% Mn to significantly lower the band gap. Under all considered substitution levels, the band edge positions and in particular the valence band edge remain in a suitable position to provide a sufficiently high overpotential for the OER that is catalyzed on these photoanode materials (see Fig. \ref{fig:CVO-MVO-bandgap-qe}a). CVO band edges are in good agreement with experimental findings for CVO ~\cite{Guo2015, Song2020}.

\begin{figure}
\centering
	\includegraphics[scale=1.0]{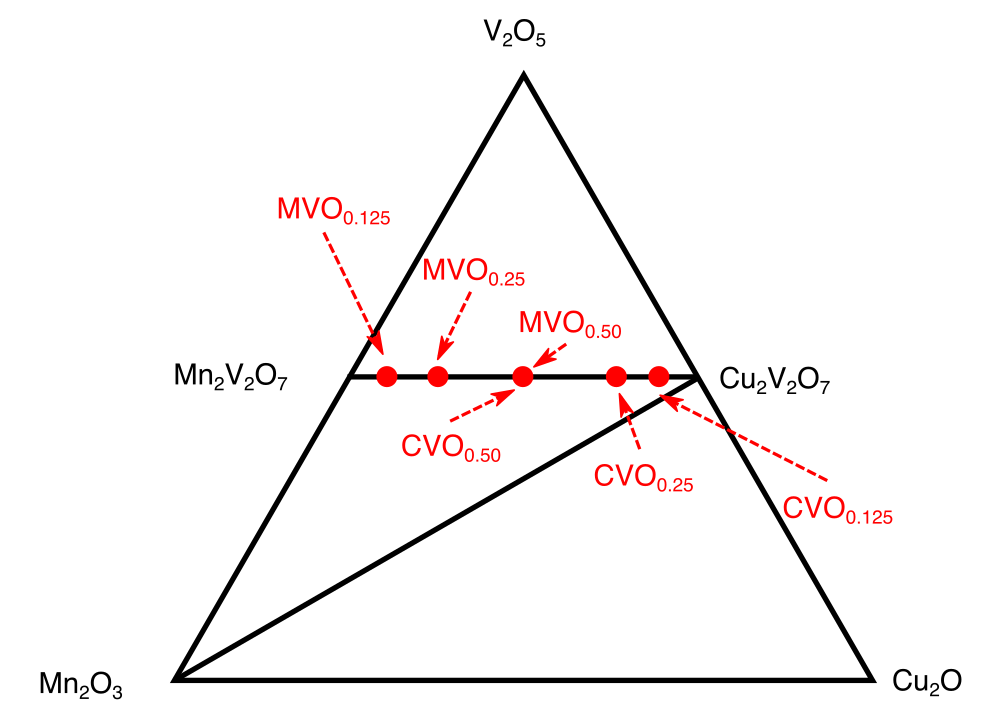}
	\caption{Phase diagram for the ternary subsystems with a chemical potential of $\mu_O$=-0.65 eV, corresponding to a temperature of 600\,K under 0.21\,atm pressure. The unstable compounds are presented in red.}
	\label{fig:phase_diagram}
\end{figure}

It was not possible to synthesize solid-solutions with Cu content higher than 10\%, as also reported in previous experimental work~\cite{Krasnenko2013Structural}. This is in agreement with our theoretical prediction that the solid-solutions are unstable with respect to the pure pyrovanadates (see ESI\dag\ Section S5 for details), the instability getting more pronounced further from the end members. Changing the experimental conditions by varying the pressure and/or temperature affects the value of the oxygen chemical potential but does not render the solid solutions stable as shown by the unstable red phases in the phase diagram (see Fig. \ref{fig:phase_diagram}). Solid solutions with low substitution levels are thus obtained as potentially metastable or entropy stabilized phases.

\section{Conclusions}
We investigated solid-solutions of $\beta$-Mn$_{2}$V$_2$O$_7$ and $\beta$-Cu$_{2}$V$_2$O$_7$ with the aim to evaluate their suitability for photoelectrocatalytic applications. Our results show that up to 20\% Cu incorporation into $\beta$-Mn$_{2}$V$_2$O$_7$ leads to an about 0.3 eV reduction in band gap, while incorporation of Mn into $\beta$-Cu$_{2}$V$_2$O$_7$ is predicted by our calculations to lead to a marked reduction in band gap of about 1 eV. In the intermediate composition range, band gaps remain roughly constant. Our calculations predict that all solid solutions are metastable, solid solutions with higher substitution levels being increasingly more difficult to form. This suggests that the catalytic activity of both $\beta$-Mn$_{2}$V$_2$O$_7$ and $\beta$-Cu$_{2}$V$_2$O$_7$ end members could be enhanced by low-level substitution with Cu and Mn respectively. Further studies are required to assess if these solid solutions are able to overcome the limitations inherent to pure $\beta$-Cu$_{2}$V$_2$O$_7$.
 
\section*{Acknowledgements}
This research was funded by the SNF Professorship Grant PP00P2\_157615. Calculations were performed on UBELIX (http://www.id.unibe.ch/hpc), the HPC cluster at the University of Bern. The H2020 Marie Curie Individual Fellowship grant, Agreement Number 753124, for M.S.

\balance

\bibliography{library.bib}

\newpage
\setcounter{page}{1}
\renewcommand{\thetable}{S\arabic{table}}  
\setcounter{table}{0}
\renewcommand{\thefigure}{S\arabic{figure}}
\setcounter{figure}{0}
\renewcommand{\thesection}{S\arabic{section}}
\setcounter{section}{0}
\renewcommand{\theequation}{S\arabic{equation}}
\setcounter{equation}{0}
\onecolumngrid

\begin{center}
\textbf{\large Supplementary information for\\\vspace{0.5 cm}
\LARGE Suitability of $\beta$-Mn$_2$V$_2$O$_7$/$\beta$-Cu$_2$V$_2$O$_7$ solid solutions for photocatalytic water-splitting\\\vspace{0.3 cm}
\large by \\\vspace{0.3cm}
Silviya Ninova, Michal Strach, Raffaella Buonsanti and Ulrich Aschauer}
\end{center}

\section{Geometry parameters}

In Table \ref{tbl:CMVO-lattice}, we report the lattice parameters of the pure compounds as well as the different solid solutions.

\begin{table}[h!]
	\caption{Lattice parameters of $\beta$-Mn$_2$V$_2$O$_7$ and $\beta$-Cu$_2$V$_2$O$_7$ after optimization with PBE+$U$, compared with experiment. Only the most stable structure is presented for the different substitution levels.}
	\label{tbl:CMVO-lattice}
	\centering
	\begin{tabular}{c|ccc|ccc|c}
\hline
 & a (\r{A]}) & b (\r{A]}) & c (\r{A]}) & $\alpha$ ($^{\circ}$) & $\beta$ ($^{\circ}$) & $\gamma$ ($^{\circ}$) & Volume (\r{A]}$^3$)\\
\hline
\multicolumn{7}{c}{$\beta$-Mn$_2$V$_2$O$_7$} \\
\hline
Expt~\citeSI{SI_Liao1996}          & 6.713 & 8.725 & 4.969 & 90.0 & 103.6 & 90.0 & 282.889 \\
MCVO$_{0.00}$              		& 6.686 & 8.825 & 5.002 & 90.0 & 102.7 & 90.0 & 287.922 \\
\hline
MCVO$_{0.00}^a$            		& 6.681 & 8.821 & 10.006 & 90.0 & 102.7 & 90.0 & 575.243 \\
MCVO$_{0.125}^a$            		& 6.673 & 8.778 & 10.017 & 90.0 & 103.3 & 90.0 & 571.078 \\
MCVO$_{0.25}^a$            		& 6.704 & 8.726 & 10.017 & 90.0 & 104.0 & 90.0 & 568.666 \\
MCVO$_{0.50}^a$            		& 6.846 & 8.637 & 10.054 & 90.0 & 106.6 & 90.0 & 573.224 \\
\hline
\multicolumn{7}{c}{$\beta$-Cu$_2$V$_2$O$_7$} \\
\hline
Expt~\citeSI{SI_Hughes1989}	& 7.689 & 8.029 & 10.106 & 90.0 & 110.3 & 90.0 & 585.345 \\
CMVO$_{0.00}$             			& 7.673 & 8.248 & 10.200 & 90.0 & 110.4 & 90.0 & 604.962 \\
\hline
CMVO$_{0.125}$             		& 7.645 & 8.308 & 10.215 & 89.8 & 110.8 & 90.4 & 606.465 \\
CMVO$_{0.25}$             			& 7.592 & 8.358 & 10.238 & 89.5 & 111.0 & 90.5 & 606.307 \\
CMVO$_{0.50}$             			& 7.524 & 8.474 & 10.320 & 90.0 & 111.9 & 90.0 & 606.307 \\
\hline
	\end{tabular}
	\begin{flushleft}
		$^a$ The cell is doubled along the \textit{c} axis, so as to ease a direct comparison with $\beta$-Cu$_2$V$_2$O$_7$. \\
	\end{flushleft}
\end{table}

\section{XRD Patterns}

\begin{figure}[h!]
	\centering
	\includegraphics[width=0.41\textwidth]{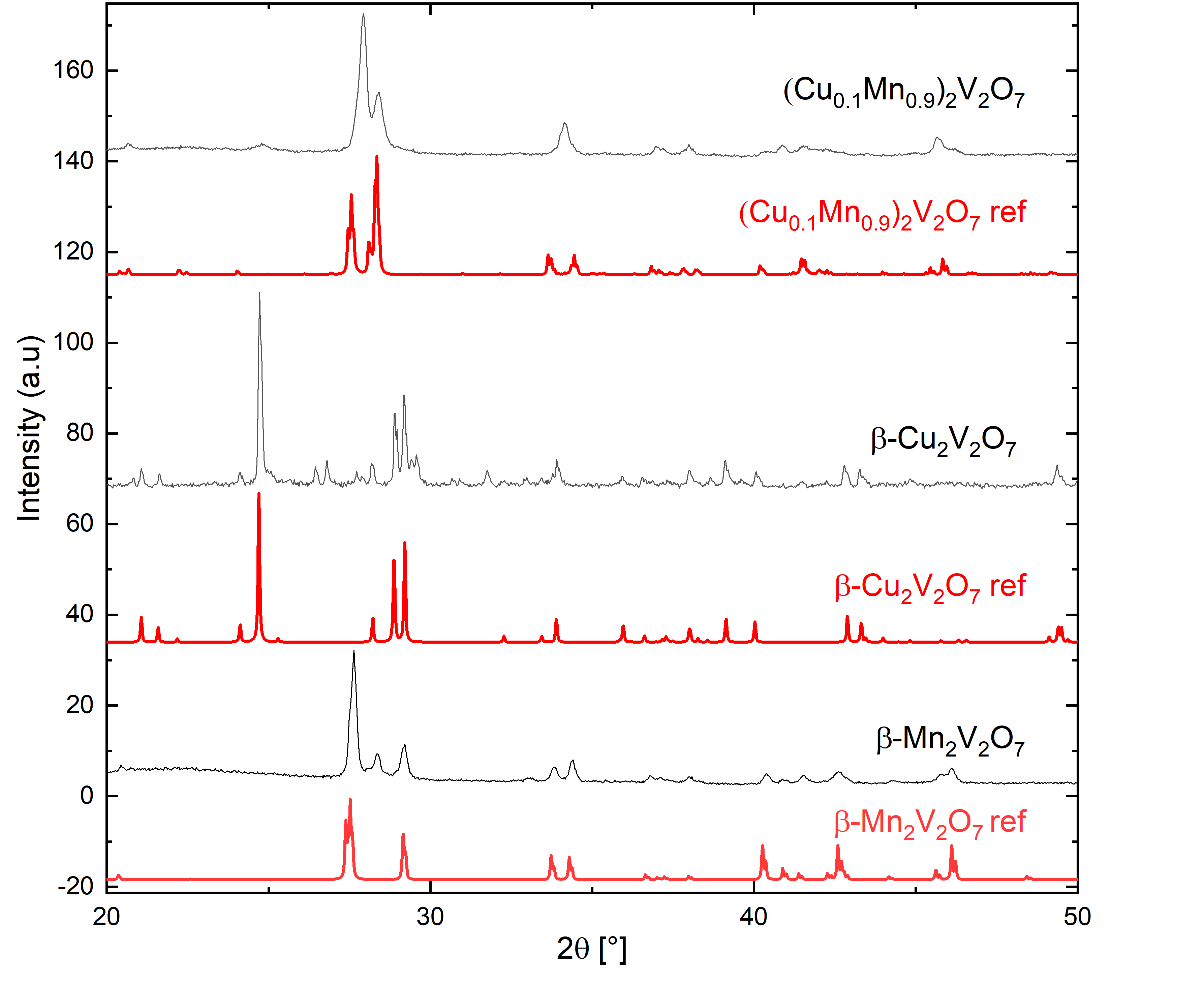}
	\caption{XRD patterns of the Cu-Mn-V-O compounds together with references (Cu$_{0.2}$Mn$_{1.8}$V$_2$O$_7$ PDF No. 04-009-0569 and pure Mn$_2$V$_2$O$_7$ PDF No. 00-052-1266).}
	\label{fig:XRD-exp}
\end{figure}

\newpage
\section{Solid-solution models}

\subsection*{$\beta$-Mn$_2$V$_2$O$_7$-derived structures}

\begin{figure}[h]
	\centering
	\includegraphics[width=\textwidth]{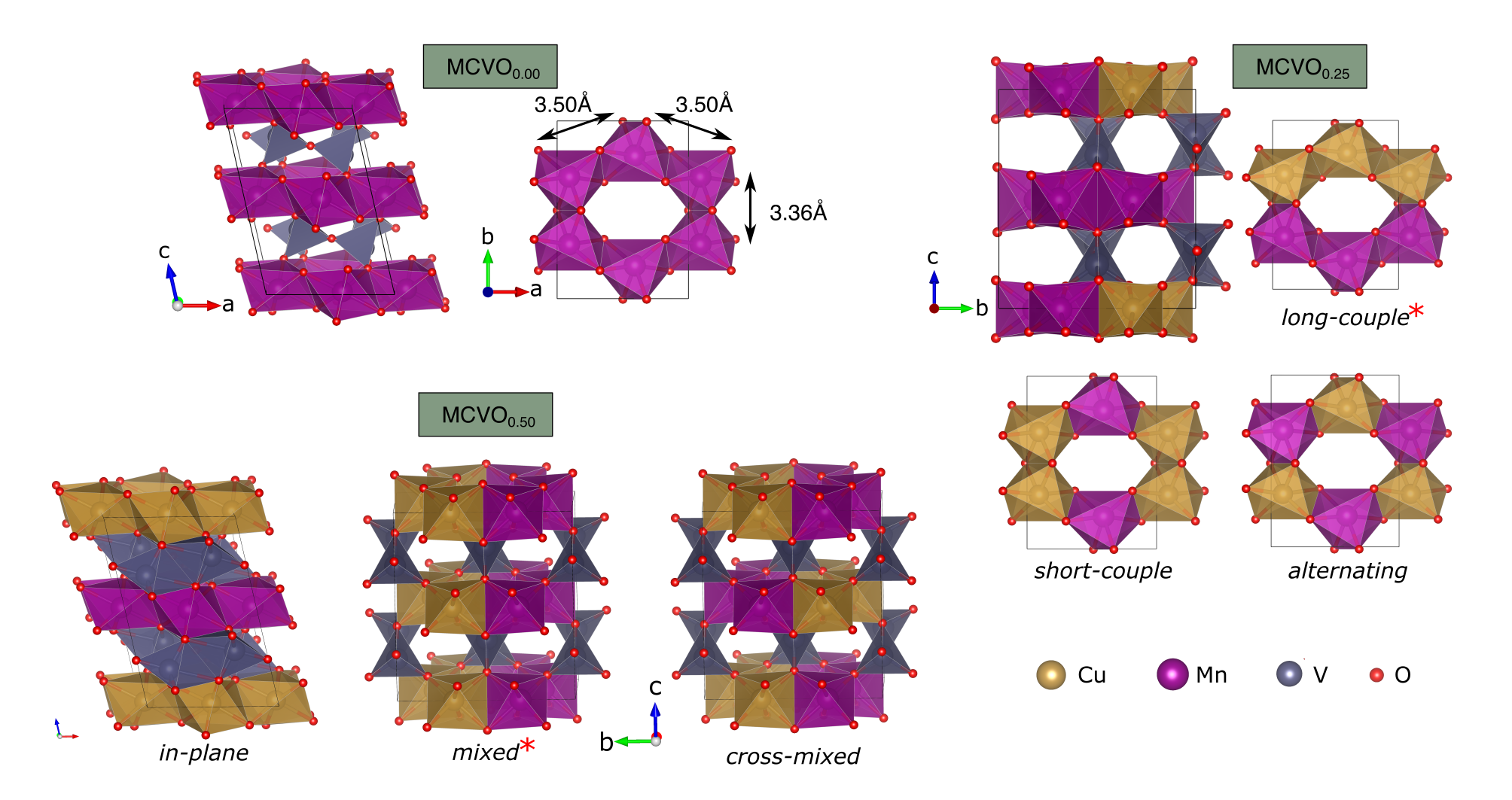}
	\caption{Several structural models are used for each stoichiometric ratio in $\beta$-Mn$_2$V$_2$O$_7$. The most stable ones are denoted with a red asterisk.}
	\label{fig:MVO-models}
\end{figure}

Figure \ref{fig:MVO-models} we show the structural models for $\beta$-Mn$_2$V$_2$O$_7$ and derived solid solutions, Table \ref{tbl:MVO-en-diff} giving energy differences between various models at x=0.25 and x=0.50.

\begin{table}[h]
	\caption{Energy differences in $\Delta$E\,(meV) per Mn (or Cu) atom for the several structural models of Cu-substituted $\beta$-Mn$_2$V$_2$O$_7$.}
	\label{tbl:MVO-en-diff}
	\begin{minipage}{.45\linewidth}
		\centering
		\begin{tabular}{ccc}  
\hline
\multicolumn{3}{c}{MCVO$_{0.25}$} \\
\hline
\hline
\multirow{2}{*}{alternating}  & AFM & 12.5 \\
\multirow{2}{*}{}             & FM  & 12.5 \\ \hline
\multirow{2}{*}{short couple} & AFM & 15.4 \\
\multirow{2}{*}{}             & FM  & 15.5 \\ \hline
\multirow{2}{*}{long couple}  & AFM &  0.3 \\
\multirow{2}{*}{}             & FM  &  0.0 \\
\hline
		\end{tabular}
	\end{minipage}
	\begin{minipage}{.45\linewidth}
		\centering
		\begin{tabular}{ccc}  
\hline
\multicolumn{3}{c}{MCVO$_{0.50}$} \\
\hline
\hline
\multirow{3}{*}{in-plane}     & AFM1 & 5.2  \\
\multirow{3}{*}{}             & AFM2 & 5.7  \\
\multirow{3}{*}{}             & FM   & 6.7  \\ \hline
\multirow{3}{*}{mixed}        & AFM1 & 0.5  \\
\multirow{3}{*}{}             & AFM2 & 0.6  \\
\multirow{3}{*}{}             & FM   & 0.0  \\ \hline
\multirow{2}{*}{cross-mixed}  & AFM  & 33.5 \\ 
\multirow{2}{*}{}             & FM   & 29.8 \\
\hline                      
		\end{tabular}
		\end{minipage}
\end{table}

\newpage
\subsection*{$\beta$-Cu$_2$V$_2$O$_7$-derived structures}

\begin{figure}[h]
	\centering
	\includegraphics[width=\textwidth]{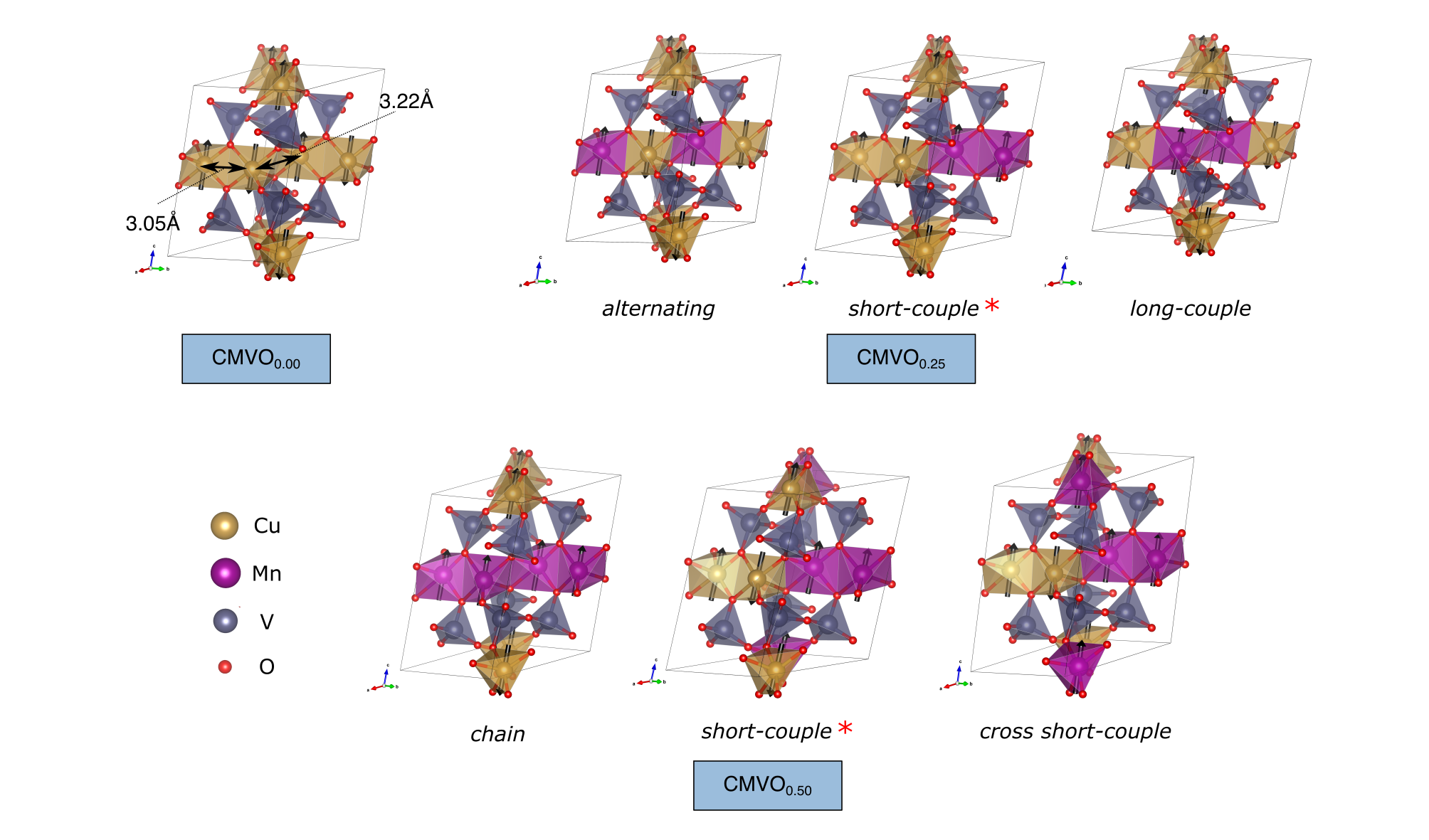}
	\caption{Several structural models are used for each stoichiometric ratio in $\beta$-Cu$_2$V$_2$O$_7$. The most stable ones are denoted with a red asterisk.}
	\label{fig:CVO-models}
\end{figure}

Figure \ref{fig:CVO-models} we show the structural models for $\beta$-Mn$_2$V$_2$O$_7$ and derived solid solutions, Table \ref{tbl:CVO-en-diff} giving energy differences between various models at x=0.25 and x=0.50.

\begin{table} [h]
	\caption{Energy differences $\Delta$E\,in (meV) per Mn (or Cu) atom for the several structural models of Mn-substituted $\beta$-Cu$_2$V$_2$O$_7$.}
	\label{tbl:CVO-en-diff}
	\begin{minipage}{.45\linewidth}
		\centering
		\begin{tabular}{ccc}  
\hline
\multicolumn{3}{c}{CMVO$_{0.25}$} \\
\hline
\hline
\multirow{2}{*}{alternating}  & AFM & 0.5 \\
\multirow{2}{*}{}             & FM  & 0.1 \\ \hline
\multirow{2}{*}{short couple} & AFM & 0.8 \\
\multirow{2}{*}{}             & FM  & 0.0 \\ \hline
\multirow{2}{*}{long couple}  & AFM & 5.6 \\
\multirow{2}{*}{}             & FM  & 4.9 \\
\hline
		\end{tabular}
	\end{minipage}
	\begin{minipage}{.45\linewidth}
		\centering
		\begin{tabular}{ccc}  
\hline
\multicolumn{3}{c}{CMVO$_{0.50}$} \\
\hline
\hline
chain              & FM & 17.4 \\
short couple       & FM &  0.0 \\
cross short-couple & FM &  6.2 \\
\hline                      
		\end{tabular}
		\end{minipage}
\end{table}

\newpage
\section{Electronic properties}

\subsection*{PBE+$U$}
The trends in band-gap change with percentage of substitution is similar with PBE+$U$ and HSE. The largest discrepancy is observed for pure $\beta$-Mn$_2$V$_2$O$_7$, where the PBE+$U$ predicts a much lower band gap of 1.41\,eV, compared to experiment and HSE (see Table \ref{tbl:hse-alpha-tuning}). The band-edges are also similar with both theoretical approaches.

\begin{figure}[h!]
	\centering
	\includegraphics[width=1.0\textwidth]{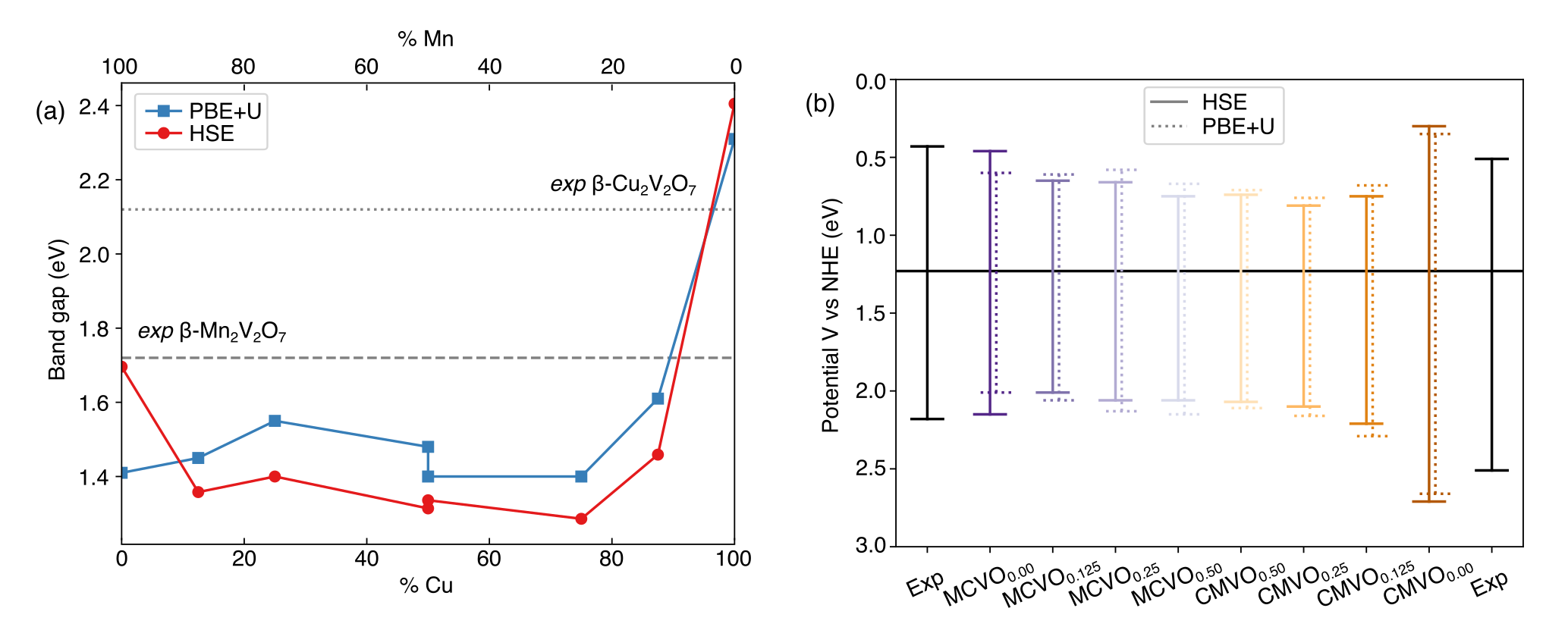}
	\caption{Comparison between the band gaps (a) and band edges (b) with PBE+$U$ and HSE.}
	\label{fig:band-gaps-edges-pbeU-hse}
\end{figure}

\begin{figure}[h!]
	\centering
	\includegraphics[width=0.7\textwidth]{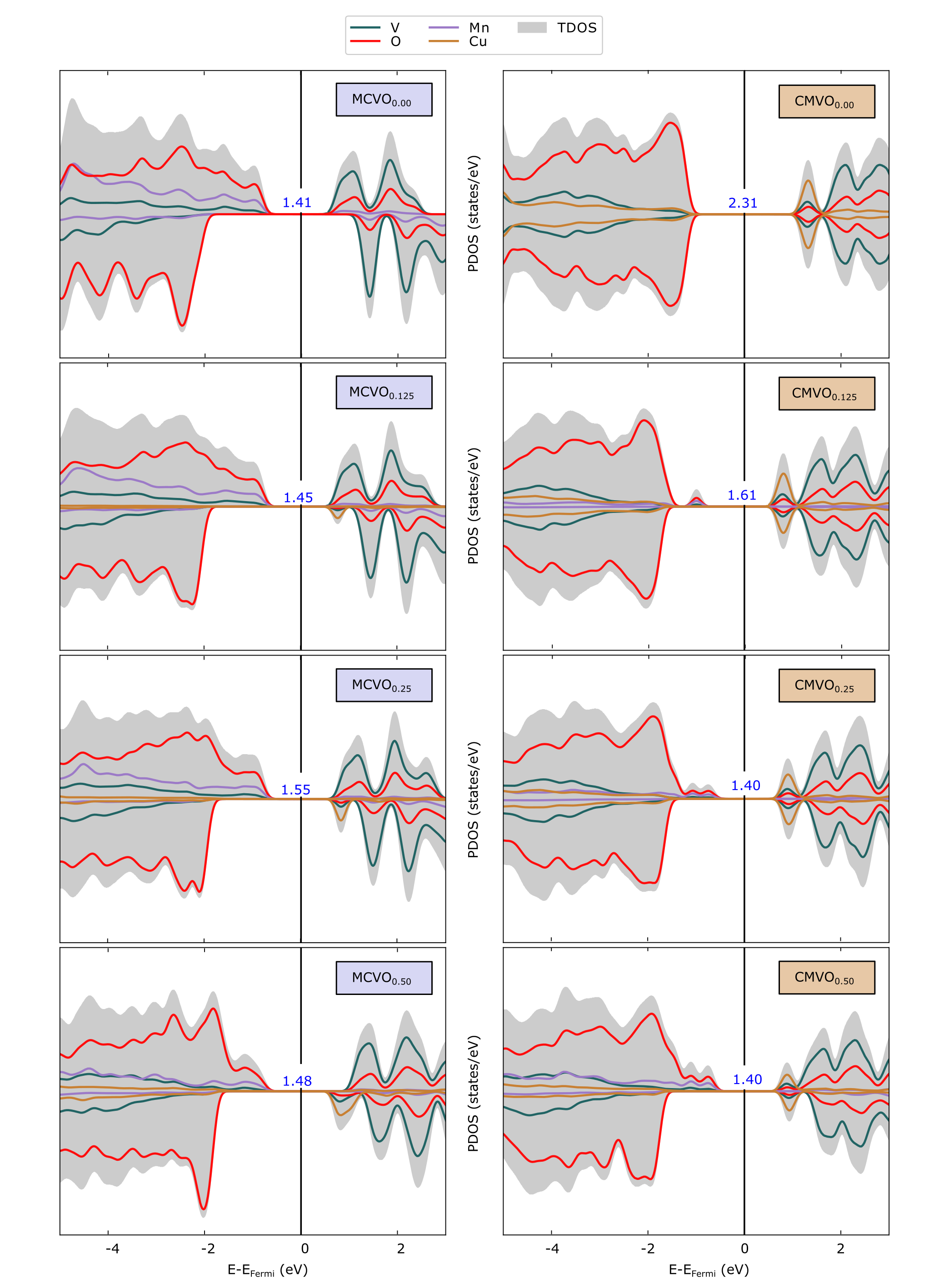}
	\caption[test]{PDOS of the most stable models for each mixing proportion, calculated at PBE+$U$ level, where the band gap (in eV) of each structure is presented in blue colour. The experimental band gaps are 1.75\,eV~\citeSI{SI_Yan2015} and 2.0 $\pm$ 0.2\,eV~\citeSI{SI_Zhou2015} for $\beta$-Mn$_2$V$_2$O$_7$ and $\beta$-Cu$_2$V$_2$O$_7$ respectively.}
	\label{fig:PDOS-pbeU}
\end{figure}

\newpage
In Figure \ref{fig:PDOS-pbeU} we show the projected density of states of the most stable structures at each substitution level computed using PBE+$U$.

\newpage
\subsection*{HSE}

In Table \ref{tbl:hse-alpha-tuning}, we compare band gaps calculated using HSE with different fractions of exact exchange ($\alpha$) to experiment. Using $\alpha=0.16$ yields the best simultaneous agreement with experiment for both $\beta$-Mn$_2$V$_2$O$_7$ and $\beta$-Cu$_2$V$_2$O$_7$.

\begin{table}[!h]
\centering
\caption{Band gap as a function of the exact exchange in the HSE hybrid functional compared to experiment.}
\label{tbl:hse-alpha-tuning}
\begin{tabular}{c|cc|cc}
\hline
 											& $\alpha=0.25$ 	& $\alpha=0.16$ 	& Experiment 	& Experiment (this work)	\\
\hline
\hline
$\beta$-Mn$_2$V$_2$O$_7$	& 2.27 					& 1.70 					& 1.75$\pm$0.1~\citeSI{SI_Yan2015}	& 1.72 \\
$\beta$-Cu$_2$V$_2$O$_7$ & 3.25 					& 2.41 					& 2.00$\pm$0.2~\citeSI{SI_Zhou2015} & 2.12 \\
\hline
\end{tabular}
\end{table}

\begin{figure}[h!]
	\centering
	\includegraphics[scale=0.8]{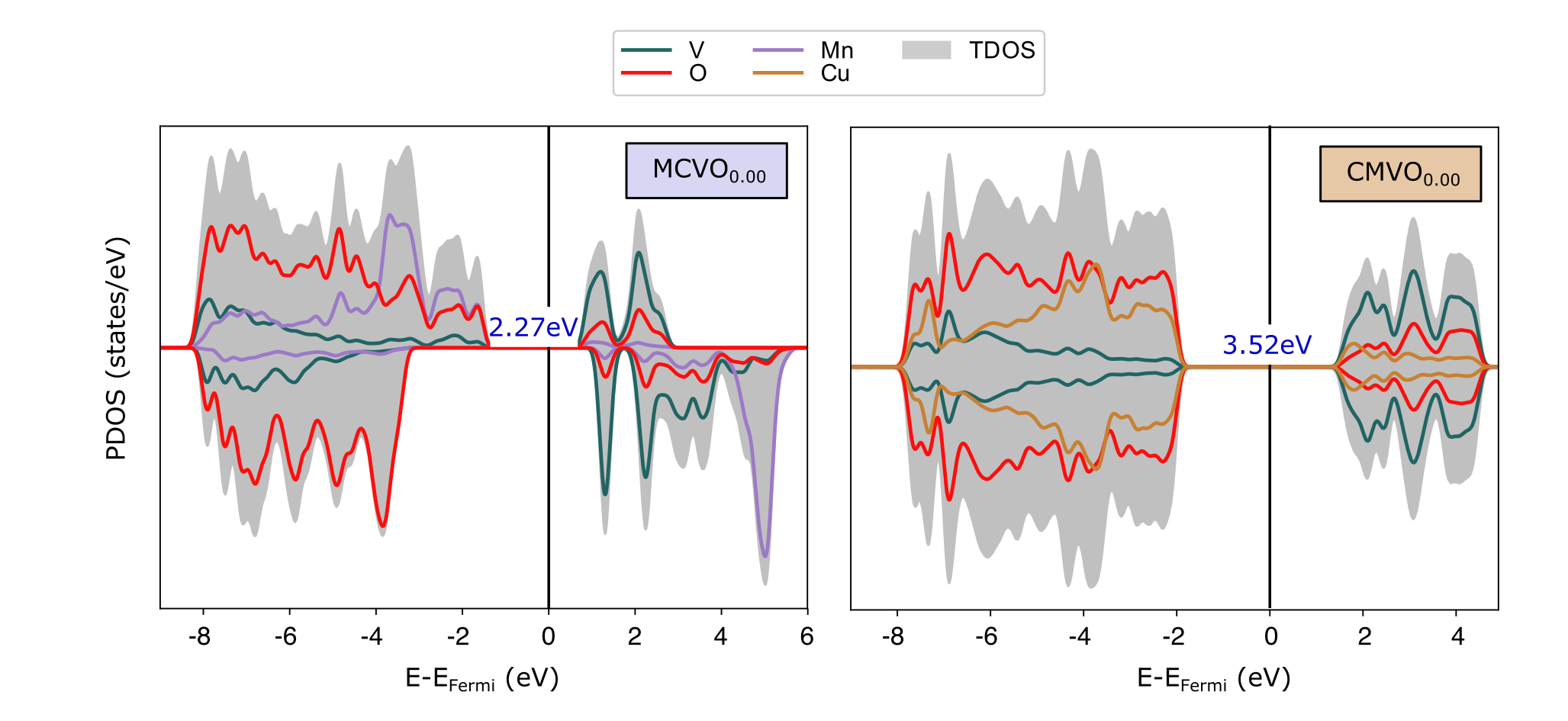}
	\caption[test]{PDOS of the pure compounds $\beta$-Mn$_2$V$_2$O$_7$ (left) and $\beta$-Cu$_2$V$_2$O$_7$ (right) as calculated using 25\% of exact exchange.}
	\label{fig:PDOS-hse-25}
\end{figure}

\begin{figure}[h!]
	\centering
	\includegraphics[scale=0.8]{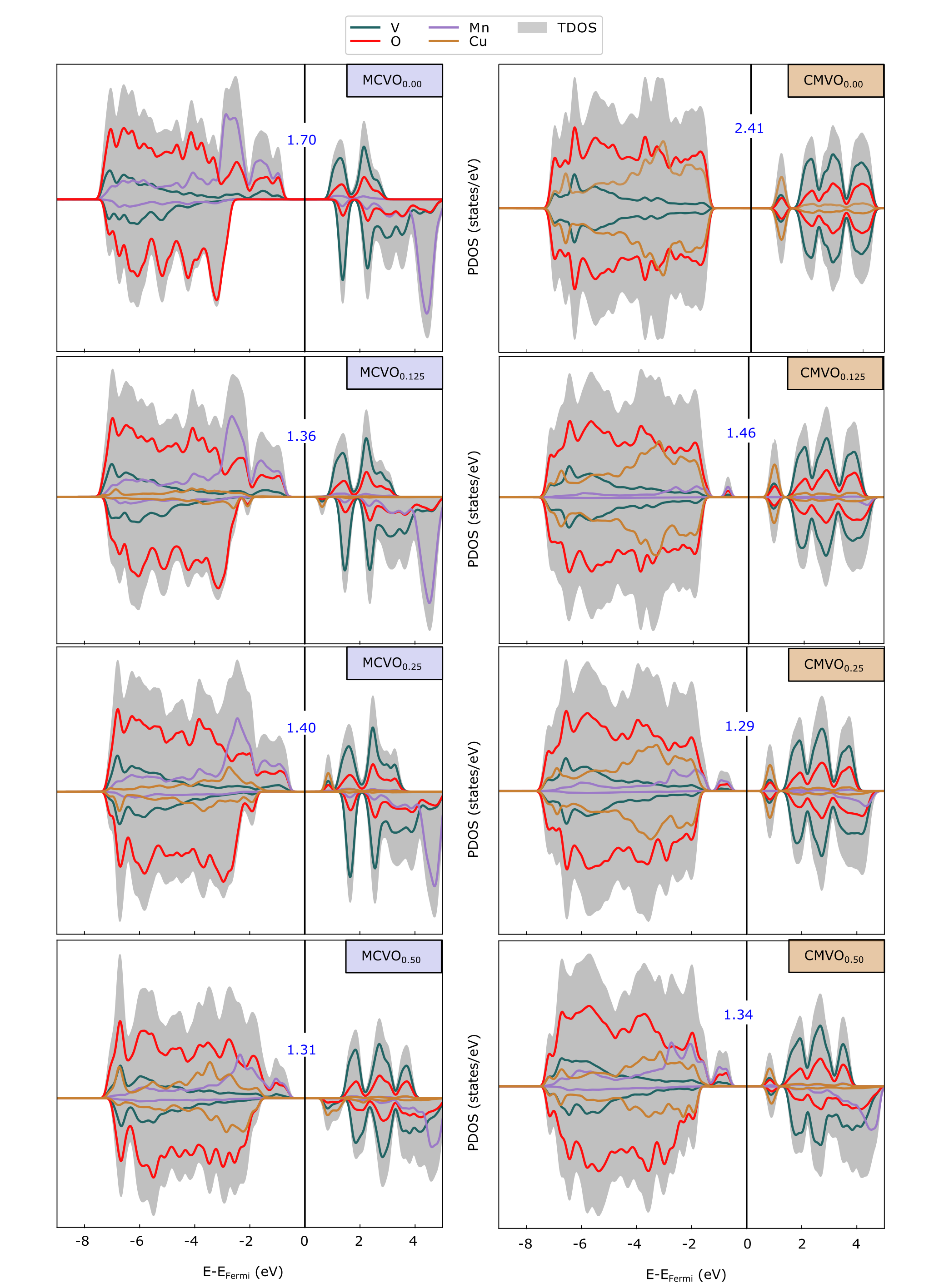}
	\caption[test]{PDOS of the pure compounds $\beta$-Mn$_2$V$_2$O$_7$ (left) and $\beta$-Cu$_2$V$_2$O$_7$ (right) and their solid-solutions as calculated using 16\% of exact exchange and the HSE functional.}
	\label{fig:PDOS-hse-16}
\end{figure}

Figure \ref{fig:PDOS-hse-25} shows the density of states with $\alpha=0.25$ for the end members $\beta$-Mn$_2$V$_2$O$_7$ and $\beta$-Cu$_2$V$_2$O$_7$, while Figure \ref{fig:PDOS-hse-16} shows the density of states for all computed solid solutions with $\alpha=0.16$.

\clearpage
\FloatBarrier
\section{Phase diagram}

We estimated the effect of different experimental conditions on the phase stability by varying the oxygen chemical potential to account for changes in partial pressures and/or temperatures. We evaluated the ternary subsystem phase diagrams following the procedure in Ong \textit{et al.} ~\citeSI{Ong2008}.

\begin{figure}[h]
	\centering
	\includegraphics[scale=0.9]{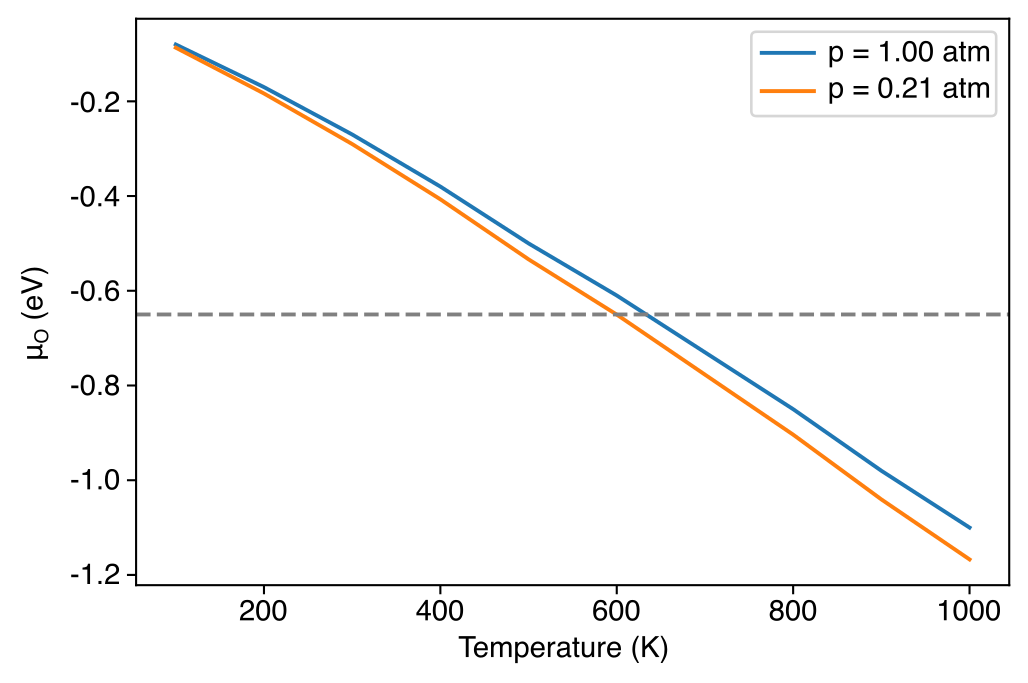}
	\caption{Temperature-pressure dependence of the oxygen chemical potential at standard $p=1atm$ and atmospheric pressure $p=0.21atm$.}
	\label{fig:pT}
\end{figure}

The dependence of the oxygen chemical potential, $\mu_{O}$ on the temperature and pressure has been evaluated as in previous works,~\citeSI{Heifets2007, Reuter2001} using the energy of the oxygen atom, $1/2E_{O_2}$, as the reference state.

\begin{equation} \label{eq:mu}
\begin{split}
\Delta \mu_{O}(T,p) & = \mu_{O}(T,p) - 1/2E_{O_2} \\
                    & = \frac{1}{2} \Delta G_{O_{2}}(T, p^0) + \frac{1}{2}kTln\left(\frac{p}{p^0} \right) + \delta \mu_{O}^0
\end{split}
\end{equation}

The first term accounts for the change in Gibbs free energy with respect to the standard conditions of $p^0=1atm$ and $T^0=298.15K$. The exact values have been evaluated following equation \ref{eq:mu-dG} with values from the thermodynamic tables~\citeSI{NIST-JANAF}.

\begin{equation} \label{eq:mu-dG}
\begin{split}
\Delta G_{O_{2}}(T, p^0) & = G_{O_{2}}(T, p^0) - G_{O_{2}}(T^0, p^0) \\
                         & = H_{O_{2}}(T, p^0) - H_{O_{2}}(T^0, p^0) - T \left( S_{O_{2}}(T, p^0) - S_{O_{2}}(T^0, p^0) \right)
\end{split}
\end{equation}

The pressure dependency is expressed in the second term in Equation \ref{eq:mu}. We considered two values for the pressure. The standard one, $p=1atm$, cancels the terms, whereas the atmospheric one, $p=0.21atm$, reduces the chemical potential by $-0.000067T$.

\begin{figure}[h!]
	\centering
	\includegraphics[scale=0.9]{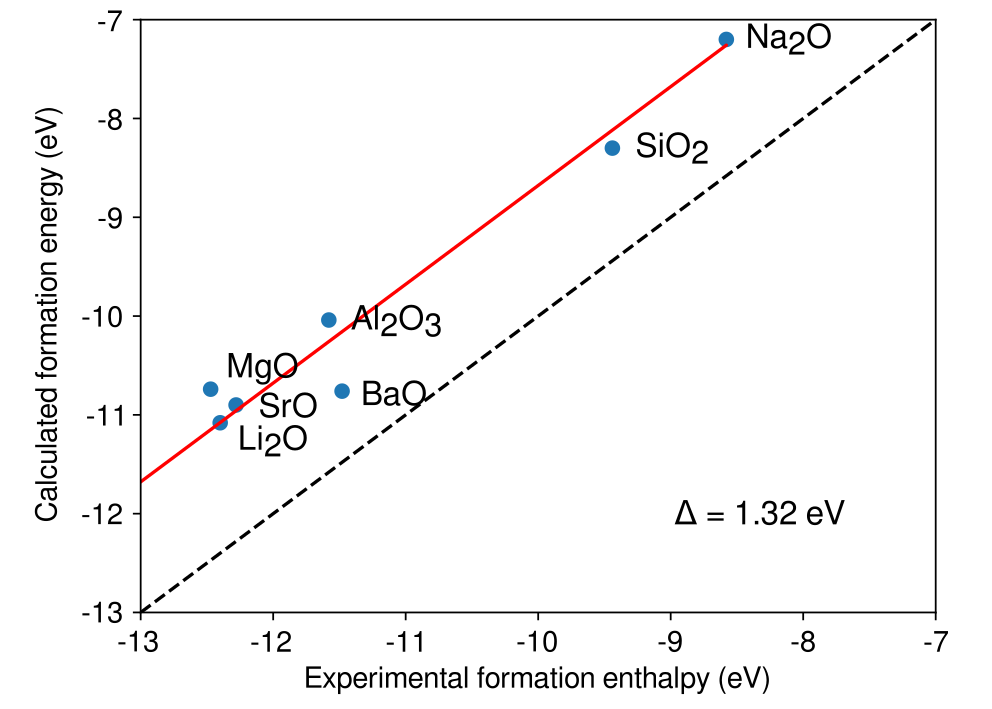}
	\caption[test]{The O$_2$ correction we find is similar to the 0.68\,eV already calculated in Wang \textit{et al.}~\citeSI{Wang2006}.}
	\label{fig:O_fit}
\end{figure}

The third term in Equation \ref{eq:mu} corrects the discrepancy between experimental data and theoretical calculations. We find a values of 1.32\,eV / 2 = 0.66\,eV (see fit in Figure  \ref{fig:O_fit}), which is in good agreement with similar calculations~\citeSI{Wang2006}.

\newpage
We applied a Hubbard parameter on all transition metals in the studied compounds. In order to correct the differences between GGA and GGA+$U$ for the calculated formation energies and the phase diagram, we determined the energy correction to be added on each atom.~\citeSI{Jain2011} The O$_2$ fit of 1.32\,eV per O$_2$ molecule (Figure \ref{fig:O_fit}) was applied in the energy. All energy numerical values are presented in Table \ref{tbl:CMVO-formation-energies}.

\begin{figure}[h!]
	\centering
	\includegraphics[width=\textwidth]{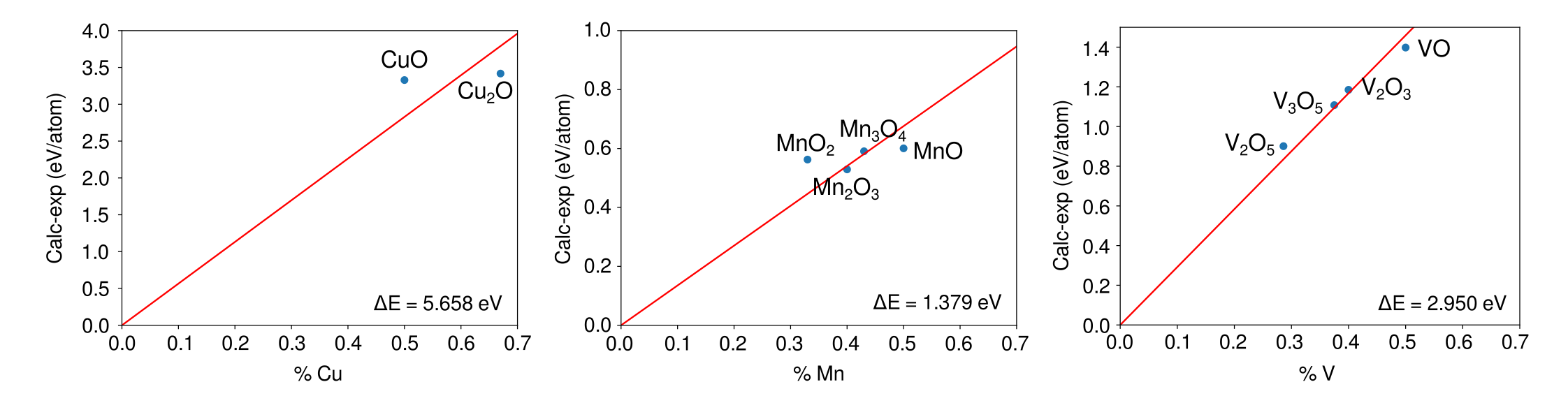}
	\caption{GGA vs. GGA+$U$ correction for the various transition metal elements for which we used a DFT+$U$ correction.}
	\label{fig:el_fit}
\end{figure}

\begin{table}[h]
	\caption[test]{Total energies used for the calculation of the phase diagram, where corrections on the metals and oxygen we applied. The following abbreviations have been used for the magnetic states: AFM - antiferromagnetic, FM - ferromagnetic, FiM - ferrimagnetic, NM - nonmagnetic. The magnetic states of the Mn-compounds have been ascribed as in ~\citeSI{Franchini2007} and references therein.}
	\label{tbl:CMVO-formation-energies}
	\centering
	\begin{tabular}{c|cc|r}
\hline
Compound 					& Space group 										& Magnetic order 	& Energy\,(eV) \\
\hline			
CuO 								& C2/c~\citeSI{Asbrink1970}				& AFM 					& -2087.42561 \\
Cu$_2$O 						& Pn$\bar{3}$m~\citeSI{Kirfel1990}		& NM 					& -3742.55480 \\
Cu$_2$O$_3$ 				& Ia3~\citeSI{Jain2013}							& FM 					& -4607.93285 \\
\hline
MnO 								& Fm$\bar{3}$m~\citeSI{Wyckoff1963}	& AFM 					& -3284.83332 \\
$\beta$-MnO$_2$ 		& P4$_2$/mnm~\citeSI{Ohama1971}		& FM  					& -3719.57165 \\
$\alpha$-Mn$_2$O$_3$& Pcab~\citeSI{Geller1971}					& FM  					& -7005.20396 \\
Mn$_3$O$_4$ 				& I4$_1$amd~\citeSI{Wyckoff1963}		& FiM 					& -10290.20148 \\
\hline
VO	 							& Fm$\bar{3}$m~\citeSI{Wyckoff1963}	& AFM					& -2394.11781 \\
V$_2$O$_3$ 					& R$\bar{3}$c~\citeSI{Dernier1970,Newnhan1962}	& AFM	& -5225.41776 \\
V$_2$O$_5$ 					& Pmmn~\citeSI{Bystroem1950}			& NM						& -6096.13000 \\
V$_3$O$_5$ 					& P2/c~\citeSI{Asbrink2002}					& FM						& -8056.15255 \\
V$_3$O$_7$	 				& C2/c~\citeSI{Waltersson1974}			& FM 					& -8926.72260 \\
\hline
MnV$_2$O$_4$				& Fd$\bar{3}$m~\citeSI{Pannunzio-Miner2009}& FM			& -8508.48720 \\
MnV$_2$O$_6$				& C2/c~\citeSI{Kimber2007}					& AFM					& -9380.14482 \\
Mn$_3$V$_2$O$_8$ 		& Cmce~\citeSI{Clemens2016}				& AFM					& -15959.42276 \\
\hline
MVO$_{0.00}$				& 	C2/m											& FM						& -51106.61482 \\
MVO$_{0.125}$				&	P1													& FM						& -49962.38385 \\
MVO$_{0.25}$				&	Pc													& FM						& -48818.32420 \\
MVO$_{0.50}$				&	Pc													& FM						& -46530.37790 \\
CVO$_{0.50}$				&	Pc/2												& AFM					& -46530.86080 \\
CVO$_{0.25}$				&	P$\bar{1}$										& AFM					& -44243.05326 \\
CVO$_{0.125}$				&	P1													& AFM					& -43099.17885 \\
CVO$_{0.00}$				&	C2/c												& AFM					& -41955.31972 \\
\hline
	\end{tabular}
\end{table}

\newpage
\bibliographystyleSI{apsrev4-1}
\bibliographySI{library}

\end{document}